\documentclass[a4paper,12pt]{article}
\usepackage[utf8]{inputenc}
\usepackage[T1]{fontenc}
\usepackage[swedish,english]{babel}
\usepackage{color}
\usepackage{amssymb}
\usepackage{amsmath}
\usepackage{ae}
\usepackage{slashed}
\usepackage{graphics}
\usepackage{graphicx}
\usepackage{epstopdf}
\usepackage{url}

\newcommand{\g}{\gamma}

\newcommand{\lb}{\bar{\lambda}}
\newcommand{\la}{\lambda}

\newcommand{\lbracket}{[\![}
\newcommand{\rbracket}{]\!]}

\newcommand{\li}{\hspace{1mm}}

\usepackage{cite}
\usepackage{epsfig}
\usepackage{times}
\usepackage[small]{caption}
\usepackage{ifthen}
\usepackage{longtable}
\usepackage{multirow}
\numberwithin{equation}{section}

\usepackage{fancyhdr}
\usepackage[colorlinks=true,linktoc=page,linkcolor=blue,citecolor=blue,filecolor=blue,urlcolor=blue]{hyperref}
\usepackage{pstricks}

\begin{document}
\selectlanguage{english}
\setcounter{secnumdepth}{3}
\frenchspacing
\pagenumbering{roman}

\null{\vspace{\stretch{1}}}
\includegraphics*[width=1.8cm]{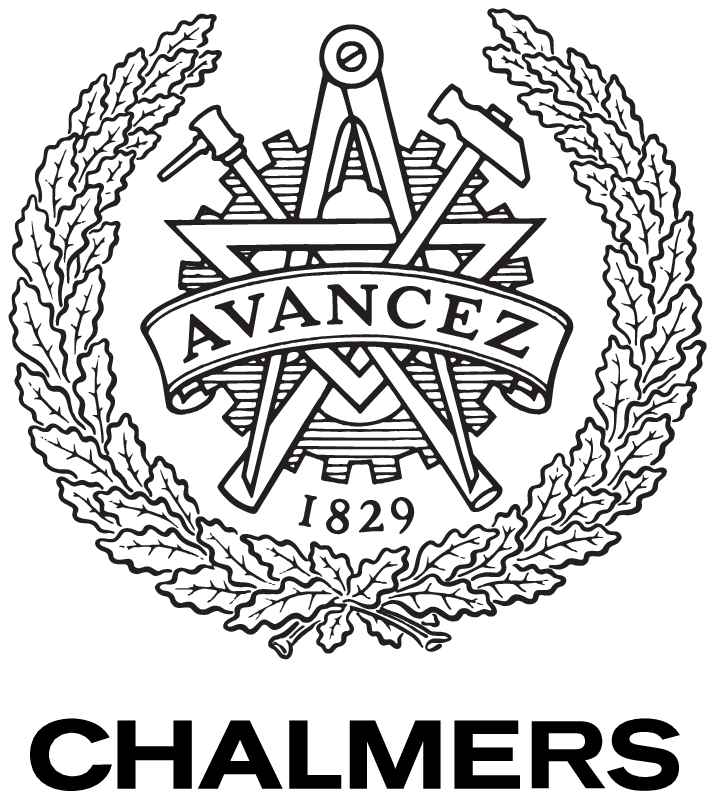}
\hfill $\begin{array}{r}\text{Gothenburg preprint} \\ \text{August, 2015}\\$\quad$\end{array}$
\vskip-3mm
\noindent\makebox[\linewidth]{\rule{\textwidth}{0.4pt}}
\vspace{\stretch{1}}
\begin{center}
{\Large Pure spinor indications of ultraviolet finiteness\\ \vspace{0.1cm} in $D=4$ maximal supergravity}\\
\vspace{\stretch{1}}
{\normalsize Anna Karlsson}\\
\vspace{\stretch{1}}
{\small Fundamental Physics \\ Chalmers University of Technology \\ SE 412 96 Gothenburg, Sweden }\\
\vspace{\stretch{1}}
\end{center}

\begin{abstract}\noindent
The ultraviolet divergences of amplitude diagrams in maximal supergravity are characterised by a first possible divergence at seven loops for the 4-point amplitude (logarithmic) and, in its absence, at eight loops. We revisit the pure spinor superfield theory results of [\href{http://arxiv.org/abs/1412.5983}{arXiv:1412.5983}], stating the absence of the divergence originating in the 4-point 7-loop amplitude as well as those of more than seven loops. The analysis, performed in terms of the one-particle irreducible loop structures giving rise to the divergences, is extended, especially with respect to the limits on the dimension for finiteness. The results correspond to those mentioned, known from other approaches, indicating an ultraviolet finiteness of maximal supergravity in $D=4$.
\end{abstract}
\vspace{\stretch{2}}
\vfill
\noindent\makebox[\linewidth]{\rule{\textwidth}{0.4pt}}
{\tiny \texttt{email: karann@chalmers.se}}
\thispagestyle{empty}
\newpage
\noindent\makebox[\linewidth]{\rule{\textwidth}{0.4pt}}
\vspace{-0.8cm}
\tableofcontents
\vspace{0.3cm}
\noindent\makebox[\linewidth]{\rule{\textwidth}{0.4pt}}
\pagenumbering{arabic}
\section{Introduction}
The ultraviolet divergences of the amplitude diagrams in maximal supergravity \cite{SA,Brink:1980az,Cremmer:1980ru,Wit:1982ig} have long been subject to investigations \cite{Grisaru:1976ua,Grisaru:1976nn,Deser:1977nt,Tomboulis:1977wd,Deser:1978br,Deser:1979sx,Berkovits:2004px,Berkovits:2006vi,Bern2,vanB,Bern25,KLR,BHS,AB,Bern3,BD,Vanhove:2010nf,Bossard:2010dq,Beisert:2010jx,BG,JB,Bern:2012uf,CK2,Karlsson:2014xva,Kallosh:2014hga,Bern:2014kca,Wang:2015jna,Wang:2015aua,Bossard:2015oxa}. However, while a divergent theory in $D=4$ initially seemed unavoidable, with increasing UV divergences the higher the number of loops present, the explicit 4-graviton calculations of \cite{Bern2,Bern25,Bern3,Bern:2012uf} showed a better UV behaviour at four loops than expected. The results of \cite{BG,JB,Vanhove:2010nf,Beisert:2010jx} then showed a first possible divergence at $L=7$ for the 4-point diagram: a logarithmic divergence also discussed in \cite{Bern:2014kca}. Interestingly, the results of \cite{Beisert:2010jx} also stated that, if the 4-point 7-loop would be absent, the 5-point 7-loop would be characterised by a slightly better UV behaviour, i.e. not divergent in $D=4$ at $L=7$.

This is all the more interesting in the light of the recent pure spinor investigations of \cite{CK2,Karlsson:2014xva}. Unlike other pure spinor approaches \cite{Berkovits:2004px,Berkovits:2006vi,AB,BG,JB}, these are performed in a field theory setting, based on the maximal supergravity action \cite{Cederwall:2009ez,Cederwall:2010tn} respecting maximal supersymmetry, and where the maximal supersymmetry is kept an inherent property throughout the investigations. Importantly, the results of \cite{Karlsson:2014xva} showed a cut-off of the possible number of loops in one-particle irreducible loop structures, and because the UV divergences only are caused by such substructures, and defined by the worst separate UV behaviour thereof, \cite{Karlsson:2014xva} effectively stated that the UV divergences only depend on the number of loops present up to $L=6$ for the 4-point amplitude, and $L=7$ for the 5-point amplitude. By the previous results stated, the possible 4-point 7-loop logarithmic divergence would be avoided, and by \cite{Beisert:2010jx} effectively all UV divergences in $D=4$, a point not explicitly made in \cite{Karlsson:2014xva}. This indicates a scenario with maximal supergravity UV finite in $D=4$, relevant for further investigations, but in part supported by arguments in e.g. \cite{Kallosh:2014hga}.

In this article\footnote{In our original analysis, also present in \cite{Karlsson:2014xva}, we mistakenly ignored the possible contribution of nonzero modes of $N_{mn}$ in the $b$-ghost. It is possible that contributions from these nonzero modes will modify our conclusions, and we are currently investigating this. We thank Nathan Berkovits for pointing this out.} we revisit the arguments in \cite{Karlsson:2014xva} and have a further look at the UV divergences. For example, the cut-off of the loop behaviour is better interpreted as a product of the integration over loop momenta, in combination with regularisation properties. The latter which in turn can be used to give an upper limit of the UV behaviour with limits on the dimension, for finiteness, corresponding to those of \cite{BG,JB,Vanhove:2010nf,Beisert:2010jx}, some of which were deduced through U-duality properties. Perhaps the corresponding processes can give some insight into what the pure spinor formalism cancellations ought to correspond to in other approaches. It at least seems to concern U-duality properties in combination with an insensitivity to certain required transformations, in terms of the loop integrations.
\\ \\
The article is organised as follows. To begin with, a brief presentation of the pure spinor formalism and the concepts connected to the amplitude diagrams formulated in a field theory setting are given. For an extensive review of the first, see \cite{Cederwall:2013vba}, and in terms of the latter, we refer to \cite{CK2,Karlsson:2014xva} or, in a brief format, \cite{Karlsson:2014xoa}.

We then proceed with a more extensive analysis of the restrictions on the one-particle irreducible loops structures observed in \cite{Karlsson:2014xva}. Especially, the effective operators can be limited further, as specified in \eqref{eq.opEff}, which brings about further limits on the UV divergences: a simple power-counting of the momenta present yield the correct UV divergences for $L\leq2$ in \eqref{eq.PestD}. This estimate, assuming an equal division of the momenta in the loop structure, clearly is a bit naive for $L>3$, where the loop configurations begin to play an important role. This is possible to take into account through further observations of the loop regularisation properties in \eqref{eq.loopEx}: a given loop, be it a part of a diagram or not, does not diverge worse when considered part of a larger loop structure than when figuring on its own, provided all contributing operator configurations are considered\footnote{Some equivalences of momenta, part of the effective operators in \eqref{eq.opEff}, effectively acting out are only valid in the presence of `true' outer legs.}. In this way the momenta actually contributing to the UV divergences can be narrowed down to present results in equivalence with \cite{Bern2,Bern25,Bern:2012uf} for $3\leq L\leq4$
and worst case scenarios in $L\leq 7$ corresponding to those of \cite{BG,JB,Vanhove:2010nf,Beisert:2010jx}, which in combination with the limits on the one-particle irreducible loop structures in \cite{Karlsson:2014xva} indicate UV finiteness for maximal supergravity in $ D=4$.

\section{Amplitudes in pure spinor field theory}
In \cite{CK2,Karlsson:2014xva} a field theory formulation of amplitude diagrams in the pure spinor formalism was set down, benefitting from the maximal supersymmetry respected by the action \cite{Cederwall:2009ez,Cederwall:2010tn}:
\begin{gather}\begin{aligned}\label{eq.SUGRA}
S_{SUGRA}=&\frac{1}{\kappa^2}\int[\mathrm{d}Z]\bigg(\frac{1}{2}\psi Q\psi+\frac{1}{6}(\la\g_{ab}\la)\Big(1-\frac{3}{2}T\psi\Big)\psi R^a\psi R^b\psi\bigg)\\[0.5cm]
R^a=&\eta^{-1}(\lb\g^{ab}\lb)\partial_b-\eta^{-2}L_{(1)}^{ab,cd}(\la\g_{bcd}D)+\hspace{4.1cm}\\
&+2\eta^{-3}L_{(2)}^{ab,cd,ef}\Big[(\la\g_{bcdei}\la)\eta_{fj}-\frac{2}{3}\eta_{f[b}(\la\g_{cde]ij}\la)\Big]N^{ij}\\
T=&8\eta^{-3}(\lb\g^{ab}\lb)(\lb r)(rr)N_{ab}
\end{aligned}\end{gather}
The pure spinor formulation originates in the linearised $D=11$ supergravity theory, which in flat superspace is possible to formulate in terms of a covariant spinor derivative acting on the 3-form $C_{\alpha\beta\gamma}$, with a structure possible to capture \cite{B1,B3,Brink:1980az,Cremmer:1980ru,Cederwall:2009ez} through the introduction of a bosonic, pure spinor of ghost number one \cite{B1,B3}:
\begin{equation}
\la_\alpha: \la\g_a\la=0.
\end{equation}
In terms of a pure spinor superfield $\psi$ with nothing but $\la^\alpha\la^\beta\la^\gamma C_{\alpha\beta\gamma}$ at $\lambda^3\lb^0r^0$ in a series expansion in the variables, the equation of motion and gauge is $Q\psi=0$ and $\delta\psi= Q\Lambda$ with
\begin{gather}\begin{aligned}
\{D_\alpha,D_\beta\}&=-2(\g^a)_{\alpha\beta}\partial_a\\
Q&=\la D +r\bar{\omega}, \quad\Rightarrow\quad Q^2=0.
\end{aligned}\end{gather}
By this construction, the component supergravity theory is retainable at ghost number zero in the minimal formalism $(x^a,\theta_\alpha,\la_\alpha)$ while the non-minimal variables $(\lb_\alpha,r_\alpha)$, counterparts to $(\la_\alpha,\theta_\alpha)$ and $\lb$ of ghost number $-1$: $(\lb\g_a\lb)=(\lb\g_ar)=0$, allow for the construction of the integral measure\footnote{The derivatives with respect to $(x,\theta,\la,\lb,r)$ are $(\partial,D,\omega,\bar{\omega},s)$, the latter three which ought only appear in the gauge invariant 2- and 0-form operators formed out of $(\la\omega,\lb\bar{\omega},\lb s)$: $(N,\bar{N},S)$.} \cite{B2}.

Importantly, $Q$ is a BRST operator and the formulation that of a BRST formalism, with an action respecting maximal super-Poincar\'e symmetry. The Batalin--Vilkovisky formalism \cite{BV,BR} therefore presents a consistent extension to interactions through the BRST operator being replaced with an action acting nonlinearly on the superfield through an antibracket \cite{Cederwall:2009ez}:
\begin{equation}
(A,B)\sim\int\frac{\delta A}{\delta\psi}\frac{\delta B}{\delta\psi}[\mathrm{d}Z],
\end{equation}
The form of which occurs due to the superfield $\psi$ containing all ghosts and antifields, effectively representing its own antifield. The subsequent formulation has the equation of motion $(S,\psi)=0$ and is correct provided $(S,S)=0$, which is how the action is set, starting from the BRST action while including the interactions stated in the superspace formulation of gravity \cite{Howe:1997he,Cremmer:1980ru,Brink:1980az,CGNN1,CGNN2,CGNT}.

\subsection{Important features of the pure spinor formalism}
The pure spinor formalism has three crucial features in BRST equivalence, gauge fixing, and integration in the presence of general regularisations. The first, BRST equivalence, originates in calculations only being performed between free, on-shell, external states (obeying $Q\psi=0$), leaving the theory invariant under
\begin{equation}
1 \leftrightarrow 1+\{Q,\chi\},
\end{equation}
provided a fermion $\chi$ of correct ghost number and dimension, with the special case of a regulator: $e^{\{Q,\chi\}}$.

Gauge fixing is, in the absence of any antifield other than $\psi$ itself, performed through a Siegel gauge \cite{Siegel} in an imitation of string theory: a $b$-ghost figuring in the free propagator as $b/p^2$ is introduced and required to fulfil
\begin{equation}\label{eq.bReq}
\{Q,b\}=\partial^2,\quad b\psi_\text{on-shell}=0.
\end{equation}
The former of these (in a BRST equivalent sense) gives $\{b,b\}=0$ and sets \cite{Karlsson:2014xva}
\begin{gather}\begin{aligned}
b=&\frac{1}{2}\eta^{-1}(\bar{\la}\g_{ab}\bar{\la})(\la\g^{ab}\g^iD)\partial_i+\\
&+\eta^{-2}L^{(1)}_{ab,cd}\Big((\la \g^{a}D)(\la \g^{bcd} D) +2(\la{\g^{abc}}_{ij}\la)N^{di}\partial^j+\\
&\qquad\qquad+\frac{2}{3}(\eta^b_p\eta^d_q-\eta^{bd}\eta_{pq})(\la\g^{apcij}\la)N_{ij}\partial^q\Big)-\\
&-\frac{1}{3}\eta^{-3}L^{(2)}_{ab,cd,ef}\Big((\la\g^{abcij}\la)(\la\g^{def}D)N_{ij}-\\
&\qquad\qquad-12\Big[(\la{\g^{abcei}}\la)\eta^{fj}-\frac{2}{3}\eta^{f[a}(\la\g^{bce]ij}\la)\Big](\la\g^{d}D)N_{ij}\Big)+
\end{aligned}\end{gather}
\begin{equation}
+\frac{4}{3}\eta^{-4}L^{(3)}_{ab,cd,ef,gh}(\la\g^{abcij}\la)\Big[(\la{\g^{defgk}}\la)\eta^{hl}-\frac{2}{3}\eta^{h[d}(\la{\g^{efg]kl}}\la)\Big]\{N_{ij},N_{kl}\},\nonumber
\end{equation}
where $L^{(p)}$ denotes
\begin{equation}\label{eq.L}
L^{(p)}_{a_0b_0,a_1b_1,\ldots,a_pb_p}=(\lb\g_{\lbracket a_0b_0}\lb)
(\lb\g_{a_1b_1}r)\ldots(\lb\g_{a_pb_p\rbracket}r),
\end{equation}
antisymmetrises the $p+1$ \emph{pairs} of indices through $\lbracket\ldots\rbracket$, and by default obeys
\begin{gather}\begin{aligned}
L^{(p)}L^{(q)}&\propto(\lb\g^{(2)}\lb)L^{(p+q)}\label{eq.L+}\\
[r\bar{\omega},\eta^{-(p+1)}L^{(p)}_{a_0b_0,\ldots,a_pb_p}\}&=2(p+2)\eta^{-(p+2)}L^{(p+1)}_{ab,a_0b_0,\ldots,a_pb_p}(\la\g^{ab}\la).
\end{aligned}\end{gather}
Moreover, is only non-zero for $L^{(p)}:p\leq15$. \cite{CK2,Karlsson:2014xva}

A non-degenerate integration measure is given by the non-minimal variables through their properties \cite{Cederwall:2009ez,Berkovits:2002uc,Anguelova:2004pg}
\begin{gather}\begin{aligned}\label{eq.TheIntegration}
\li[\mathrm{d}\la]\la_{\alpha_1}\ldots\la_{\alpha_7}&\sim\star{T_{\alpha_1\ldots\alpha_7}}^{\beta_1\ldots\beta_{23}}\mathrm{d}\la_{\beta_1}\ldots\mathrm{d}\la_{\beta_{23}}\\
[\mathrm{d}\lb]\lb_{\alpha_1}\ldots\lb_{\alpha_7}&\sim\star{T_{\alpha_1\ldots\alpha_7}}^{\beta_1\ldots\beta_{23}}\mathrm{d}\lb_{\beta_1}\ldots \mathrm{d}\lb_{\beta_{23}}\\
[\mathrm{d}r]&\sim\lb_{\alpha_1}\ldots\lb_{\alpha_7}\star{\bar{T}^{\alpha_1\ldots\alpha_7}}_{\hspace{1cm}\beta_1\ldots\beta_{23}}\frac{\partial}{\partial r_{\beta_1}}\ldots \frac{\partial}{\partial r_{\beta_{23}}},
\end{aligned}\end{gather}
where $T$ projects into $(02003)$, but general regularisations are necessary due to the bosonic $(\la,\lb)$: in the limit of infinity and on singular subspaces. The first is remedied by a regulator $e^{-(r\theta+\la\lb)}$ which also furnishes the required $(\theta,r)$ for the fermionic integrations to capture the correct dynamics. The latter is caused by scalars $\xi=(\la\lb)$ and $\eta=(\la\g^{ab}\la)(\lb\g_{ab}\lb)\sim\xi^2\sigma^2$ present in the theory, where $\sigma$ refers to the 2-form subspace. Only a limited negative power of these $(\xi^{-22},\sigma^{-11})$ \cite{Cederwall:2009ez} can be part of a convergent integrand, often calling for a second generalised regularisation: \cite{Berkovits:2006vi}
\begin{equation}
O_\text{reg}(\la,\bar{\la}) =\int[\mathrm{d}f][\mathrm{d}\bar{f}][\mathrm{d}g][\mathrm{d}\bar{g}]e^{-\{Q,\bar{f}g\}}e^{i\varepsilon\{Q,gW+\bar{f}V\}}O(\la,\bar{\la}).
\end{equation}
Effectively, the introduction of a new set of variables $(f^\alpha,\bar{f}_\alpha,g^\alpha,\bar{g}_\alpha)$ counterpart to $(\la^\alpha,\lb_\alpha,\theta^\alpha,r_\alpha)$, a $Q$ extended akin to from the minimal to the non-minimal formalism, and a regulator acting on $(\la,\lb)$ through gauge invariant operators \cite{CK2} in combination with the integration, regularises the operator $O$ (in a heat-kernel way) by what was initially allowed for in terms of the singular subspaces. This procedure can be performed any number of times and so any integrand built from convergent operators (effectively all) can be regularised, but the procedure severely complicates analyses, best performed prior to the generalised regularisations. Provided the analysed entity presents a convergent integrand, the results are BRST equivalent. Otherwise, results vanishing due to the variables subject to a change under generalised regularisations (i.e. all) are void, representing $0\times\infty$ with a possible non-zero result at the regularisation of the divergence.

\subsection{Properties of amplitude diagrams}\label{ss.propAd}
The action \eqref{eq.SUGRA} describes the vertices present in the theory: the 3-point and 4-point vertices, the first with two $R$-operators acting out on two separate fields and the second in addition containing $T$ acting on a third field. In both cases, different configurations of on which fields the operators act are equivalent, and apart from these entities, the tree amplitude diagrams are constructible from the propagator and external states, with the addition of an overall integration.

At the formation of loops, however, the propagator is too local in $(\la,\lb)$ to describe loops on its own, necessitating the introduction of something like a generalised regularisation. The solution, inspired from string theory, consists of recognising the loop momenta as variables in the loop structure: $D\rightarrow \sum_ID^I$ etc. In addition, the loop regularisation \cite{Berkovits:2006vi} includes a regulator with exponent
\begin{equation}\label{eq.loopreg}
k\big((\la D)S+(\la\g_{ab} D) S^{ab}-N\bar{N}-N_{ab}\bar{N}^{ab}\big),\quad k>0.
\end{equation}
and an integration over the new variables $(\partial^I,D^I,N^I,\bar{N}^I,S^I)$ for each loop $I$, in total yielding a formulation where loops structures can be formed and analysed. Effectively, each loop integration demand $(\bar{N}^{23},S^{23})$ from the regulator, since the operators in the loops do not contain those entities. Due to the form of the regulator, this moreover satisfies $[\mathrm{d}N^I]$ and brings down $\la^{23}D^{23}$, antisymmetrised with any $r$ or $D$ due to (illustrated for a regularised $r$)
\begin{equation}\label{eq.laD23asym}
\{(\la D\lb s)^{23},e^{\{Q,\chi\}}re^{-\{Q,\chi\}}\}\propto[(\la D)^{23},e^{\{Q,\chi\}}re^{-\{Q,\chi\}}](\lb s)^{23}.
\end{equation}
In fact, all of the $D$s go into $[\mathrm{d}D^I]$ by the loop derivatives (of loop $I$) equivalently being positioned on one propagator (not shared between loops) at the loop integration\footnote{For each loop, there is at least one propagator carrying only the loop momenta of that loop. This is a given and any propagator can equivalently be considered to fill this function (though not any constellation thereof with $L>2$) but the concept is useful at an analysis of the loop structures. Equivalently, that is where loop integration takes place, and for further use the propagator will be termed `integration propagator'.}. As that is where $D^{23}$ is brought down, and $D$ is fermionic with $32$ degrees of freedom, anything but $D^9$ from the loop structure and $D^{23}$ from the regulator yields zero, all immediately claimed by $[\mathrm{d}D^I]$.

During all of this, factors of $(\la,\lb)$ are brought down from the regulator through
\begin{gather}\begin{aligned}\label{eq.dNbNS}
\li[\mathrm{d}N] \la_{\alpha_1}\ldots\la_{\alpha_{16}}&\sim{M_{\alpha_1\ldots\alpha_{16}}}^{a_1b_1\ldots a_{22}b_{22}}\mathrm{d}N_{a_1b_1}\ldots \mathrm{d}N_{a_{22}b_{22}}\mathrm{d}N\\
[\mathrm{d}\bar{N}] \lb_{\alpha_1}\ldots\lb_{\alpha_{16}}&\sim{M_{\alpha_1\ldots\alpha_{16}}}^{a_1b_1\ldots a_{22}b_{22}}\mathrm{d}\bar{N}_{a_1b_1}\ldots \mathrm{d}\bar{N}_{a_{22}b_{22}}\mathrm{d}\bar{N}\\
[\mathrm{d}S] &\sim\lb_{\alpha_1}\ldots\lb_{\alpha_{16}}{\bar{M}^{\alpha_1\ldots\alpha_{16}}}_{\hspace{1.1cm}a_1b_1\ldots a_{22}b_{22}}\mathrm{d}S^{a_1b_1}\ldots \mathrm{d}S^{a_{22}b_{22}}\mathrm{d}S
\end{aligned}\end{gather}
with $M$ projecting into $1\mathrm{x}(05006)+1\mathrm{x}(06004)+1\mathrm{x}(07002)+1\mathrm{x}(08000)$ through the overlap between 16 pure spinors and the antisymmetrisation of $22$ 2-form entities, compare to \cite{B2}. The loop integration then corresponds to
\begin{equation}\label{eq.IntLDd}
\la^7D^{-9}[\mathrm{d}\partial^I], 
\end{equation}
where the remaining integration may cause divergences in terms of $\partial^2$ (unpaired momenta yield zero), and e.g. gives rise to the UV divergences. In particular, the loop regularisation effectively acts only on $r$.

What remains is an analysis of the singularity properties with respect to $(\la,\lb)$. The loop regularisation takes care of a number of singularities through bringing down $\la$, in combination with the $\sigma$ properties of $M$ etc. Invariably, however, some structures remain divergent and in need of (further) regularisation for a consistent analysis. Important to remember in this, is that
\begin{itemize}
\item[---] convergent entities (by integrand standards) such as e.g. low-loop structures and how two operators act on one another can be examined consistently without regularisation.
\item[---] vanishing, divergent expressions are primarily avoided by a regularisation of $r$ through the loop regularisation, for $L$ effectively
\begin{equation}\label{eq.rTransf}
e^{\{Q,\chi\}}r_\alpha e^{-\{Q,\chi\}}:\quad r_\alpha \rightarrow r_\alpha+k(\g_{ab}\lb)_\alpha(\la\g^{ab}D)
\end{equation}
but, when the entity is non-zero with $r^x:x\neq0$, the expression is BRST equivalently examined with $r$ remaining unregularised. E.g. $L^{(p)}$ is nonzero only up to $p=15$, so that $r$s from $L$ equivalently are regularised down to that number, and no further \cite{Karlsson:2014xva}. The full regularisation provides an entity as convergent (or divergent) as provided by the term with $r^0$.
\end{itemize}
With a restriction to conclusions drawn in these settings, the analysis may proceed.

\section{One-particle irreducible loop structures}
In the pure spinor field theory setting, UV divergences can equivalently be analysed in terms of one-particle irreducible loop structures. This because the divergences occur in terms of the loop integrations over the loop momenta corresponding to $x$, and loop momenta are not shared between loop structures merely connected by a single propagator. Only the momenta (originally) part of a one-particle irreducible loop structure, inside the loops, share in the loop momenta and, provided a non-zero result, affect the end properties. The overall divergence of an amplitude diagram is set by the constituent one-particle irreducible loop structure diverging the most.

\subsection{Effective operators}
An important feature of loop structures is that momenta acting along the loop propagators do not act out of the loop(s) or onto anything inside the loop(s) unless forced to. Furthermore, if remaining in a loop, they must be integrated out by the loop integration in order not to constitute total derivatives, falling under the first point just listed. Considering the loop integration properties mentioned after \eqref{eq.loopreg}, this constrains the parts of the field theory operators (the propagator and the operators in the 3- and 4-point vertices) that are necessary to consider with respect to non-vanishing results. If $\partial$ cannot form $\partial^2$ it must, just like $N$, be forced to act out of the one-particle irreducible loop structure, or onto another entity inside it. In particular, the only cause for this to happen to the bosonic momenta ($D$ is another matter) is by $b^2=0$, as discussed in \cite{Karlsson:2014xva}. However, the vanishing of certain momenta can be specified further than what is done there; further restrictions which also are valid in the \cite{Karlsson:2014xva} discussion on the case of maximally supersymmetric Yang--Mills theory.

To begin with, consider a 4-point vertex as part of a one-particle irreducible loop structure. If it constitutes an outer vertex, $T$ can equivalently be taken to act into the loop structure, and otherwise it certainly does, resulting in an $N$ in the loop(s) which invariably will constitute a total derivative. 4-point vertices therefore are not part of non-zero one-particle irreducible loop structures.

Next, consider a $b$-ghost (containing two derivatives) acting across a vertex, i.e. from one propagator to another:
\begin{gather}\begin{aligned}\label{fig.bAct}
\begin{picture}(10,27)(25,0)
\put(-20,0){\line(1,0){100}}
\put(30,0){\line(0,1){17}}
\put(-5,7){$\underrightarrow{b}$}\put(55,3){$b$}\put(17,20){$b$ or $\psi_{\text{on-shell}}$}
\end{picture}
\end{aligned}\end{gather}
This is a process equivalently examined with the $r$s remaining behind: a consideration necessary for later regularisations to be valid. Also, any $R$s may be considered to have acted past the $b$s next to the vertices. If both derivatives in the $b$ acting across the vertex then act onto the same state, the result is zero either by $b^2$ or \eqref{eq.bReq}, as the considered entity is convergent. In this way, $b$ is split onto two propagators, one of which might be an outer leg. Once split, $b^2=0$ does not occur unless both derivatives sidle up again. Anyhow, for one of the derivatives to act out, this must occur next to an outer leg (by the initial configuration). There, it is equivalent to choose which of the two derivatives acts across the vertex first, and it is only the other one that is forced out. Consequently, if a derivative in $b$ would yield zero by staying in the loop, it can be (equivalently) chosen to stay in the loop, with a zero result. Hence, any part of the operators containing an $N$ gives a vanishing result, and can be disregarded at examinations.

In addition, $(\lb\g^{mn}\lb)\partial_m$ cannot pair up into $\partial^2$, so that any term proportional to it gives a vanishing expression for the same reasons as just stated for $N$. The reason for its absence is due to the properties of $\lb$, see appendix \ref{app.Lb}. With $\lb\g_{mn}\lb$ implicit, the operators, apart from $\partial^m$ and in the absence of $N$, contain $\partial$ and $D$:
\begin{equation}\label{eq.listofD1}
(\la\g^{mn}\g^iD)\partial_i\quad(\la\g^m D),
\end{equation}
and in the presence of regularised $r$s, by \eqref{eq.rTransf} also:
\begin{equation}\begin{array}{rl}
(\lb\g_{\lbracket ab}\lb)(\lb\g_{cd\rbracket}r): &4(\lb\g_{\lbracket ab}\lb)(\lb{\g_{c}}^m\lb)(\la\g_{d\rbracket m} D),
\end{array}\end{equation}
effectively presenting entities 
\begin{equation}\label{eq.listofD2}
(\la\g^{mn}D)\quad(\la\g^{mj} D): j\neq n,
\end{equation}
not acting on the derivatives of the operators originally on the same propagator, as that expression in a BRST equivalent sense contains $[r,D]$ (antisymmetrised). However, out of these $D$s, the only $\partial$s possible to form are
\begin{gather}\begin{aligned}
&\{(\la\g^m D),(\la\g^{nj} D)\}\propto(\la\g^{mn}\la)\partial^j\\
&\{(\la\g^{mi} D),(\la\g^{nj} D)\}\propto (\la\g^{mnijs}\la)\partial_s,
\end{aligned}\end{gather}
and $\partial_m$ paired up with either of these four existing $\partial$s gives zero.

Due to the effective absence of $N$ and $(\lb\g^{mn}\lb)\partial_m$ inside the one-particle irreducible loop structures, the only operators yielding non-zero results are
\begin{gather}\begin{aligned}\label{eq.opEff}
b^{\text{eff.}}_{\text{loop}}&=\frac{1}{2}\eta^{-1}(\bar{\la}\g_{ab}\bar{\la})(\la\g^{iab}D)\partial_i+\eta^{-2}L_{ab,cd}^{(1)}(\la\g^aD)(\la\g^{bcd}D)\\
(R^a)^{\text{eff.}}_{\text{loop}}&=-\eta^{-2}L_{(1)}^{ab,cd}(\la\g_{bcd}D).
\end{aligned}\end{gather}
That said, the parts effectively yielding zero are not irrelevant. They are still present up until loop integration, taking care of properties such as $b^2=0$.

An interesting feature in connection to this discussion on equivalent treatments, especially in relation to $b$ being split while acting across an outer vertex, concerns\footnote{The two effective parts of $b$ will be denoted by $b_n$, with $n$ stating the power of $r$ in the absence of regularisation. Sometimes, this is also used for $R$. Also note the term $j$-point: $j$ outer legs connecting to states beyond the one-particle irreducible loop structure.} $b_0$. Consider the same situation as in \eqref{fig.bAct}. At an outer vertex, $\partial$ in $b_0$ might be equivalently taken to act in. If that yields a zero due to the restrictions on $D$ for loop integration, this means that the term drops out. If it does not, $\partial$ can equivalently be regarded as acting out, with $D$ in. This presents different classes of equivalence. For example, in the 4-point 1-loop amplitude ($r^x$: $x\leq8$), the only non-zero element is $(b_1)^4(R_1)^4$ with $D^3$ (one from each of the three\footnote{In $j$-point, $L$-loop one-particle irreducible loop structures, $j$ propagators are outer (caused by the presence of the outer legs) provided $L>1$. For $L=1$, this number instead is $j-1$.} outer propagators) acting out due to $b^2=0$.

\subsection{Loop structure constraints}
Effectively, the one-particle irreducible loop structures consist only of 3-point vertices, propagators and outer legs. With $L$ loops and $j$ outer legs, there are
\begin{gather}\begin{aligned}\label{eq.noPV}
&3(L-1)+j &&\text{propagators} \\
&2(L-1)+j &&\text{vertices}
\end{aligned}\end{gather}
$j$ of the latter which are outer, where one of the $R$s equivalently might be taken to act out, thereby not contributing to the divergences of the loop structure. This structure furthermore needs to provide $D^{9}$ to each $[\mathrm{d}D^I]$ for a non-zero result; $D$s not originating in the regularisation of $r$.

The last is a both obvious and subtle feature; at most $D^{23}$ can be claimed from the regulator, effectively also by regularised $r$s, as both provide 0- and 2-form $\la D$, of which no more than 24 can be antisymmetrised with a non-zero result. Furhtermore, the $\la^{24}D^{24}$ in question cannot be paired with the $\la D$ in \eqref{eq.listofD1} to form $\la^{32}D^{32}$; the irreducible representations do not match. Rests then the statement above. However, this concerns a much larger structure than what has been dealt with up until this point, so it is best to check for its convergence.

The entity $\la^{24}D^{24}$ is part of a regulator, not a singular operator (possibly) subject to regularisation, and does not encode any singularities. The eight operators $\la D$, do. In the presence of fully regularised $r$s, the effective operators behave like 
\begin{gather}\begin{aligned}\label{eq.opDiv}
&b_0\sim\big(\lb\xi^{-1},\sigma^{-1}\big)&&b_1\sim\big(\lb\xi^{-1},\sigma^{-2}\big)\\[-0.3cm]
(\xi,\sigma):\quad&\\[-0.3cm]
&(R_1)^2_\text{inner}\sim\big(\lb^2\xi^{-2},\sigma^{-3}\big)&&(R_1)_\text{outer}\sim\big(\xi^{0},\sigma^{-1}\big),
\end{aligned}\end{gather}
where it is taken into consideration that the two operators ($R^a$,$R^b$) in a 3-point vertex are connected by $(\la\g_{ab}\la)$, as dictated by \eqref{eq.SUGRA}, which sits in the vertex (part of the loop structure, unlike the $R$s acting out from the outer vertices). The most striking feature is that while a regularisation of $r$ brings about $\la\lb\sim\xi$ so that there is no difference in behaviour between $b_0$ and $b_1$, the same is not true for $\sigma$. $\lb$ pairs up into the required irreducible representation, $\la$ does not. It is more strongly coupled to the $D$s, and so in general remains to be analysed in that setting.

Effectively and equivalently, three or (if $b:b_1$) four of the eight $\la D$ originate in $RbR$ on the integration propagator. With respect to $\xi$, the worst possible behaviour is $\xi^{-6}$, clearly convergent ($\xi^x:x>-23$). With respect to $\sigma$, it is possible include the properties of the integration over the momenta. The effects of $[\mathrm{d}\bar{N}][\mathrm{d}S]$ in this respect cancel each other, and the by $[\mathrm{d}N]$ and $\la^{23}D^{23}$ remaining $\la^7\sim\sigma^2$. A worst behaviour then is set by eight $\la D$s from $(R^2b_0b^5)$ with $\sigma^{-13}\sigma^2\sim\sigma^{-11}$ or from $(R^2b_1b^4)$ with $\sigma^{-12}\sigma^2\sim\sigma^{-10}$ (under consideration: one loop), also convergent ($\sigma^x:x>-12$). Consequently, it is equivalent to treat the entity without considering further regularisation.

With the conclusion shown to be valid, it is possible to return to the requirement of $D^{9L}$ (original) going into the loop integrations, claimed from the operators in the loop structure. The $b$-ghost can at most provide two (on outer legs one) $D$, and the same goes for the vertices. However, there is a subtlety with respect to the derivatives at $L>2$: no more than $(\la\g^mD)^2$ can be antisymmetrised and go into $D^9$, so that the inner $b$s at most can contribute with $[3(L-1)+2L]$ $D$s. This yields a requirement valid for any (sub-) one-particle irreducible loop structure: \cite{Karlsson:2014xva}
\begin{gather}\begin{aligned}\label{eq.LD9L}
&L=1:&1+2j&\geq9\\[-0.35cm]
&&&&\quad\Rightarrow\quad j\geq 4.\\[-0.25cm]
&L\geq2: &9L-7+2j&\geq9L\\
\end{aligned}\end{gather}

This brings us to the last result of \cite{Karlsson:2014xva}: the limit on $L$ by the shape of the effective operators. Recall \eqref{eq.laD23asym}. When considering a loop, $RbR$ (by equivalence) is present on the integration propagator. Equivalently, so are $\la^8D^8$ from the operators (some from $RbR$), where all of the $D$s and the $RbR$ $r$s are antisymmetrised\footnote{By considering the $RbR$ to be unregularised, an additional $\sigma^2$ is present in $Rb_0R$ and $\sigma^3$ in $Rb_1R$. However, with $j\geq4$, the number of $\la D$s originating in $R$ is at least four, decreasing the worst estimate by $\sigma^2$, and so the entity is convergent by integrand standards, in the presence of loop integration.}. Also equivalently, the $r$s on the propagator (originating in $RbR$) are regularised, as in \eqref{eq.rTransf}, with both its parts $r+\la\lb D$ fully antisymmetrised with the other $(r,D)$ in the expression. At the integration, $\la^{23} D^{23}$ is brought down and antisymmetrised with these entities, as in \eqref{eq.laD23asym}, at which point there is a regularised expression (convergent as examined right above) with non-zero contributions only from the parts not proportional to $\la^{24}D^{24}$ formed out of 0- and 2-forms, i.e. originating in the regulator. This draws on the observation right above, confirming $D^9$ (original) to be required for the loop integration. Here, instead, the conclusion is that while $r$ is regularised, the integration gives a vanishing result by $\la\lb D$ from that regularisation, provided the original $r$ sits on the integration propagator.

Now this is interesting, because since the effective $R\propto r$, there is at least $r^2$ on the integration propagator of a loop. When $L>7$, there is $r^{x}: x>15$ on the integration propagators of the loops, and in total the expression vanishes. Moreover, at $j=4$ the requirement of $D^{9L}$ specifies this further. In such a one-particle irreducible loop structure, all but one of the $D$s possible to obtain from the structure are required ($L>1$), as specified in \eqref{eq.LD9L}. In specific, $(2L-1)$ $b_1$s must be present on the inner propagators, of which there are $3(L-1)$. Only $(L-2)$ may carry $b_0$, and so at least two of the integration propagators must carry $b_1$ instead of $b_0$. A non-zero result then requires $2L+2\leq15\Rightarrow L\leq6$. In total,
\begin{gather}\begin{aligned}\label{eq.limOpIs}
&L\leq6 & j=4\\
&L\leq 7 & j\geq5
\end{aligned}\end{gather}
is the one-particle irreducible loop structure requirements for non-vanishing results. Loop structures with a higher number of loops are of course allowed for, but only as a product of multiple one-particle irreducible loop structures. Furthermore, as the UV divergences are set by the individual divergences of the one-particle irreducible loop structures, no amplitude diagram diverges more than the ones in \eqref{eq.limOpIs}.

Important to note, is that it is not the regularisation of $r$ that `fails' in this. The $r$s on the propagators are regularised, the $r^x: x>15$ are equivalently zero, and the remaining terms are set to zero by the loop integration. If $L>7$, the loop integrations give a vanishing result.

For a confirmation of the validity of the results, it is possible to look at the convergence of the entities under consideration at the different points where conclusions are drawn. When considering the first results yielding zero, those are given by $(R^2)^8$ and $(R^2)^7b^2$ on the integration propagators: convergent entities when regularised, the first at worst behaving like ($\xi^{-16}$,$\sigma^{-8}$) and the second like ($\xi^{-16}$,$\sigma^{-11}$). Some of the expansions of the regularised $r$s on the propagators are therefore cut off with respect to the power of $r$, equivalently also for any $L>8$. When analysing such an integral, the expression on the integration propagator also is convergent, as analysed before, but with an extra 0- or 2-form $\la D$ which gives a vanishing result at the integration. Since the criteria specified at the end of section \ref{ss.propAd} have been met, the conclusions are valid regardless of the actual $(\la,\lb)$ subspace singularities of the amplitude diagrams.

\section{The UV divergences}\label{s.UVdivs}
A first, naive estimate of the worst possible UV divergences of a one-particle irreducible loop structure is provided by two procedures: a look at the divergences in the absence of regularisation of $r$, and a power-counting of what might combine into $\partial^2$ inside it, when $r$ is regularised.

The first is set by the number of free $\partial$s in the structure, the same as the number of $D$s remaining inside after loop integration. At a worst estimate, they can combine into a number of $\partial^2$s described by
\begin{equation}\label{eq.unReg}
L>1:[L/2+j-5],
\end{equation}
appropriately rounded off, i.e. to the closest (lower) integer. At $L=1$, $\partial$s on outer legs can equivalently be taken to act out, prohibiting $\partial^2$ from forming.

The second can be termed in $r$. To begin with, regularisation demands $r^x$: $x>15$, by \eqref{eq.opEff} and \eqref{eq.noPV}: $7(L-1)+2j>15$, and the variable only needs to be regularised down to $r^{15}$. Note that regularisation is absent only for
\begin{gather}\begin{aligned}\label{eq.noReg}
&L=1 &j\leq7\\
&L=2&j=4.
\end{aligned}\end{gather}
At regularisation, it is equivalent to consider $\partial^2\sim D^2 \sim r^4$ with $R\sim r^2$ and $b\sim r^3$, taking into account that $b$ on outer propagators lose at least $D\sim r$ to the outside, $D^{9L}$ is claimed by the loop integration and full $\partial^2$s must be possible to form. The worst estimate then gives a number of $\partial^2$s:
\begin{gather}\begin{aligned}\label{eq.nReg}
&2L-9+j\quad &L=1\\
&2L-8+j &L>1,
\end{aligned}\end{gather}
representing a positive number in the absence of \eqref{eq.noReg}.

Because the momenta $\partial^2$ present in the one-particle irreducible loop structure at loop integration(s) are given by the propagators $b/\partial^2$ and the $m$ entities formed out of the operators $(b,R)$, the UV divergences from the $L$ $[\mathrm{d}\partial]$s appear with a requirement for finiteness according to
\begin{equation}\label{eq.GlimitD}
LD-6(L-1)-2j+2m<0,
\end{equation}
which with $m$ as specified above (both regularised and non-regularised) restricts finiteness to
\begin{gather}\begin{aligned}\label{eq.PestD}
&D<8 &L=1\\
&D<2-\frac{10}{L} &L\geq2 && \Rightarrow &&D<7 &&L=2,
\end{aligned}\end{gather}
where the $L\leq2$ properties are set, \emph{definitely}, by the properties at \eqref{eq.noReg}. However, the estimate for $L\geq3$ presumes the momenta to be shared equally between the loops, with no restrictions; an unlikely situation, and moreover proven wrong by \cite{Bern2,Bern25,Bern:2012uf} with $L=3: D<6$ and $L=4:D<11/2$.

\subsection{A careful look at the UV divergences}
At the discussion on BRST equivalent examinations in connection to \eqref{eq.rTransf}, we noted that an entity can be examined equivalently in the presence of $r^{15}$ provided it is non-zero. For a loop containing $r^x$, this means that the expression for it is
\begin{equation}\label{eq.loopEx}
\propto\Big(r^x +r^{x-1}\la\lb D + \ldots +(\la\lb D)^x\Big) \sim (\partial^2)^y,
\end{equation}
where the last statement does not refer to the $\partial^2$ formed out of the $r$s, but the UV behaviour of the loop in terms of momenta formed out of $(R,b)$. It is set either by $r^x$ or, if $x>15$, by $r^{15}$. An important point here is that an addition of $r^z$ outside the loop does not alter the UV behaviour of the loop. In the extreme: at an addition of $r^{15}$ outside the loop, equivalently chosen not to be regularised, the last term in the expansion \eqref{eq.loopEx} is picked out as the only non-zero contribution. Importantly, this \emph{does not alter} the UV behaviour of the loop. Plenty of $r$s get regularised, but they do not increase the number of $\partial^2$s formed. Especially, if the loop is part of a one-particle irreducible loop structure and there are regularised $r$s (with respect to the discussed last term) that neither are shared by another loop, nor possible to fit into the $\partial^2$s formed (originally), those $r$s cannot go into the $\partial^2$s responsible for the UV divergences. In an estimate as in \eqref{eq.nReg}, these ought to be removed.

In particular, the structures that require further investigation have $L>2$ and by \eqref{eq.noReg} \emph{require} regularisation. The removal of $r$s, as described, from \eqref{eq.nReg} is valid provided the unregularised limit of \eqref{eq.unReg} is respected, which will be implicit in most of the discussion from here on.

Also possible to note is that further (general) regularisation changes the upper limits on the UV divergences as deduced in this section (\ref{s.UVdivs}) no more for divergent entities than for the convergent ones; i.e. not at all. The effective shape of the regulator can be observed in \cite{Berkovits:2006vi}, but with respect to the exponent, it only contains the $(\partial,D,r)$ variables in constellations of $g\bar{f}\la D$, $r s$ and some additional $s$. The last derivative is however equivalently only claimed by $[\mathrm{d}s]$ from the loop regulator, so the two last types of expressions add no further $r$ to the expressions. The first one moreover only has $\la D$ in 0- and 2-form constellations, same as the loop regulator, so that it cannot be claimed by $[\mathrm{d}D]$ for the reasons already stated in connection to regularised $r$s. Neither can it act on any $D$, because the fermion $g$ must be claimed by $[\mathrm{d} g]$, effectively bringing about the same situation as in \eqref{eq.laD23asym}.

\subsubsection*{Illustrative example of `irrelevant' $r$s}
Consider the two loop (sub-) structure $L=2$, $j=5$:
\begin{equation}\nonumber
\begin{picture}(185,50)(-25,0)
\put(20,20){\circle{40}}
\put(20,0){\line(0,1){40}}
\put(37,31){\line(2,1){12}}\put(37,9){\line(2,-1){12}}\put(40,20){\line(1,0){12}}
\put(3,31){\line(-2,1){12}}\put(3,9){\line(-2,-1){12}}
\put(-24,2){\ldots}\put(-24,37){\ldots}
\put(4,40){$\mathfrak{1}$}\put(30,40){$\mathfrak{2}$}
\put(80,30){$\mathfrak{1}$: $L=1$, $j=4$}\put(80,10){$\mathfrak{2}$: $L=1$, $j=5$}
\end{picture}\end{equation}
which by \eqref{eq.nReg} has a maximum of $\partial^2$ formed by the operators inside. The first loop ($\mathfrak{1}$) clearly does not cause any $\partial^2$ to form. The same does not go for the second ($\mathfrak{2}$). Had it been a one-particle irreducible loop structure, no $\partial^2$ would have been possible to form by that all but one $\partial$ (part only of $b^\text{eff.}$) equivalently act out on the outer legs. However, in the loop structure specified above, with loop integration considered to take place on the propagators denoted by ($\mathfrak{1}$,$\mathfrak{2}$), one propagator is shared between the two loops. $b_0\propto D\partial$ then carries momenta split between the loops, e.g. $D^\mathfrak{1} \partial^{\mathfrak{2}}$, where the $\partial$ cannot equivalently be assumed to act out. When it acts into the loop $\mathfrak{2}$, it can couple up with the $\partial$ in $b_0$ on the integration propagator, giving $\partial^2$. Moreover, this gives a non-zero result as the $D^9$ requirement is met with $b_1$ on the three outer legs --- moreover the only effective contribution from those outer propagators as discussed right after \eqref{eq.opEff}.

Now, the example above is a bit specific as it is a 2-loop structure with two parts $\sim(r^{15},r^{17})$, both sporting the same divergence. However, with the structure extended as indicated by the dots (with at least one loop), the `$r$'s of loop $\mathfrak{1}$-$\mathfrak{2}$ which are not shared with the other loop(s) may equivalently be regularised, except for $RbR$ on the integration propagators\footnote{It is desirable to keep the restrictions on the loop structures observed by $RbR$ on the integration propagators, and so is is necessary to keep that configuration in the analysis of the loops.}. This is $R^5$ and $b^5$: in total $r^{10}$ of which only $r^4$ may go into $\partial^2$. However, in this setting there is also the $D$ present to be considered: $D^{18}$ (three inner propagators with $b\propto D^2$), 18 of which are required for the $[\mathrm{d}D]$s. Effectively, there is a loss of $r^{6+0}$ (any extra $D$ would have been added) in the estimate of \eqref{eq.PestD}.

Moreover, if the extended structure (not in any way indicated) require the $r$ on the shared propagator for its maximal formation of $\partial^2$, and we only consider the part of the $\mathfrak{1}$-$\mathfrak{2}$-loop $\propto r^{17}$, there is a contradiction. \emph{Both} loops cannot use the same, regularised $r$ --- it ought not be counted twice. One of the formations of $\partial^2$ in reality falls short by one $r$, effectively an entire $\partial^2$.

\subsubsection*{Principles of the extended analysis}
There are a few rules to observe in this, best listed in general. However, keep the illustrative example above in mind. The first point made is that
\begin{itemize}
\item[---] for a general use, i.e. when part of a one-particle irreducible loop structure is considered, \eqref{eq.PestD} and updated limits for $L\geq3$ can be used with \eqref{eq.GlimitD} to obtain the maximal number of $\partial^2$s possibly formed in a sub-loop structure. For the specific behaviour, depending on the structure (as we soon shall illustrate), further analysis is required. E.g. a 1-loop structure at most has $[(j-4)/2]$ $\partial^2$s, possible to reduce down to none only if no propagator is shared: the behaviour depends on the number of $\partial$s not equivalently acting out of the loop.
\end{itemize}
Important to remember in the continued analysis is that the set behaviour is not restricted to certain configurations of operator terms in the loop(s), such as ($\mathfrak{2}$) and ($\mathfrak{1}$,$\mathfrak{2}$). The expansion in \eqref{eq.loopEx} merely states the worst behaviour, consequently also valid for configurations which in a 1- or 2-loop setting does not give rise to the specified formation of $\partial^2$. For example, in ($\mathfrak{2}$) two $b_0$ were required for $\partial^2$ to form. In ($\mathfrak{1}$,$\mathfrak{2}$) this is not true. By regularisation, $\partial^2$ is formed in the absence of $b_0$. In the amplitude diagram above, this term exists also, but the principle is crucial in higher loop structures where \eqref{eq.unReg} falls short of capturing the full divergence. In this way,
\begin{itemize}
\item[---]the formation of $\partial^2$ in the substructures set an upper limit on the total $\partial^2$.
\end{itemize}
Often, however, the actual situation can be specified further. There is typically an over-counting of $r$, which can be corrected in two ways:
\begin{enumerate}
\item The sub-loop structures can be analysed in terms of whether or not all $r$s and $D$s by regularisation go into $\partial^2$ and $[\mathrm{d}D]$. If loop-specific variables $\sim r$ cannot be fitted into these structures, they are equivalently lost in the counting of \eqref{eq.nReg}, by equivalently being subject to further regularisation in the larger structures.
\item In the presence of many shared propagators, the actual division of $r$ between the connected loop structures is of relevance, as those $r$s equivalently can be considered to be regularised. Such an $r$ cannot belong to more than one loop, with the $\partial$ of $b_0$ in this setting equivalently represented by $rD$. $D$s on the other hand are naturally split by $b^2=0$.
\end{enumerate}
In the presence of many shared propagators, the latter most often is the most efficient approach. When $r$ is shared between two subset loop structures in a regularised expression, each can be considered to at most diverge by what is set by the separate $j$-point $L$-loop structures. However, that estimate counts the $r$, equivalently regularised, that are shared between the loops \emph{twice}, which cannot be. Either one of the substructures yield no $\partial^2$ in the process or the $r^x$ shared removes the worth of $x/4$ --- rounded towards the (closest) \emph{larger} integer --- in terms of $\partial^2$ from the estimate, whichever removes the \emph{least} $\partial^2$s. E.g. two 2-loop structures sharing $r^5$ and sporting $j=(5,6)$ in this at most show a $(\partial^2)^2$, by that the 5-point 2-loop is considered to yield what $r$s are shared to the more divergent entity. On the other hand, if both loop structures are 6-point (or more), $(\partial^2)^2$ should be removed from the subset estimate.

However, it is important to note that free (inner) $\partial$s (not required in terms of $D$ for $[\mathrm{d}D^I]$) still may combine into $\partial^2$, limited by \eqref{eq.unReg} both in terms of the part and the whole, in the `part absence' of regularisation just discussed. When that is an issue, it is equally practical to note that there is an over-counting of $D$ in the loop substructures, by $R$ effectively being divided between the loops, on the shared propagators, in the same way as just discussed in terms of $r$. Since $D\sim r$, this also limits the the combinations in the different substructures, e.g. a $\partial^2$ formed out of the free $\partial$s in one substructure effectively removes some power of $r$ from the connecting loop(s).

By these procedures, it is possible to analyse the actual $L\leq3$ limits on finiteness. Moreover,
\begin{itemize}
\item[---] the overall limits are set by the minimal $j$.
\end{itemize}
The fact that a higher number of outer legs by no means cause a worse divergence is e.g. possible to discern in an iterative manner from $j_{min}$. If the UV behaviour at $j$ is known, the worst possible UV behaviour of $j+1$ is obtainable through $j\rightarrow j+1$ with the extra outer leg equivalently added to the integration propagator (equivalently put anywhere in the diagram). As such the addition of the leg at most brings $\sim r^4$ into that loop (one $R$ and $b$ less a $D$ or $\partial$, forced out by $b^2=0$) which at most might add one extra $\partial^2$, countered by the $1/p^2$ of the additional propagator.

We will now proceed with a re-evaluation of \eqref{eq.PestD} for $3\leq L\leq4$. Subsequent to that, we will use the described principles and the further observations made to provide new estimates of the worst behaviour sported by 4- and 5-point one-particle irreducible loop structures of $L\leq 7$.

\subsubsection*{The 4-point $L=3$}
\begin{figure}[tbp]
\begin{center}
\begin{picture}(95,75)(-30,-20)
\put(20,20){\circle{40}}
\put(-25,40){a)}
\put(14,0){\line(0,1){39}}\put(26,0){\line(0,1){39}}
\put(40,20){\line(1,0){12}}\put(0,20){\line(-1,0){12}}
\put(-5,-8){\rotatebox[origin=c]{90}{\Big\{}}\put(23,-8){\rotatebox[origin=c]{90}{\Big\{}}
\put(-2,-18){$+1$}\put(26,-18){$+1$}
\put(-2,37){$\mathfrak{1}$}\put(16,44){$\mathfrak{2}$}\put(39,37){$\mathfrak{3}$}
\put(16,-17){$\big\uparrow$}
\end{picture}
\begin{picture}(95,75)(-30,-20)
\put(-25,40){b)}
\put(20,20){\circle{40}}
\put(20,20){\line(0,1){20}}\put(20,20){\line(-1,-1){14}}\put(20,20){\line(1,-1){14}}
\put(1,28){\line(-2,1){10}}
\put(-1,38){$\mathfrak{a}$}\put(42,27){$\mathfrak{b}$}\put(26,16){$\mathfrak{c}$}\put(-8,16){$\mathfrak{1}$}
\put(4,-8){\rotatebox[origin=c]{77}{\bigg\{}}
\put(9,-17){$\mathfrak{2}$-$\mathfrak{3}$}
\put(60,35){$+2$}
\put(58,32){\line(1,0){20}}\put(58,45){\line(1,0){20}}\put(58,32){\line(0,1){13}}\put(78,32){\line(0,1){13}}
\end{picture}
\\
\begin{picture}(82,80)(-27,-19)
\put(-25,40){c)}
\put(20,20){\circle{40}}
\put(10,2){\line(0,1){35}}\put(30,2){\line(0,1){35}}\put(20,-1){\line(0,1){41}}
\put(40,20){\line(1,0){12}}\put(0,20){\line(-1,0){12}}
\put(-7,-9){\rotatebox[origin=c]{90}{\Big\{}}\put(25,-9){\rotatebox[origin=c]{90}{\Big\{}}
\put(-4,-18){$+1$}\put(28,-18){$+1$}
\put(-3,36){$\mathfrak{1}$}\put(10,44){$\mathfrak{2}$}\put(25,44){$\mathfrak{3}$}\put(37,35){$\mathfrak{4}$}
\put(17,-3){$\bullet$}\put(17,37){$\bullet$}
\end{picture}
\begin{picture}(82,71)(-27,-19)
\put(-25,40){d)}
\put(20,20){\circle{40}}
\put(12,1){\line(0,1){37}}\put(23,20){\line(1,0){17}}\put(23,0){\line(0,1){40}}
\put(0,20){\line(-1,0){12}}
\put(-7,-9){\rotatebox[origin=c]{90}{\Big\{}}\put(20,-10){\rotatebox[origin=c]{90}{\Big\{}}
\put(-4,-18){$+1$}\put(25,-19){$+2$}
\put(-3,36){$\mathfrak{1}$}\put(12,43){$\mathfrak{2}$}\put(34,38){$\mathfrak{3}$}\put(34,0){$\mathfrak{4}$}
\put(21,-3){$\bullet$}\put(21,37){$\bullet$}
\put(14,-18){$\big\uparrow$}
\end{picture}
\begin{picture}(102,87)(-20,-20)
\put(-15,40){e)}
\put(20,20){\circle{40}}
\put(14,0){\line(0,1){39}}\put(26,0){\line(0,1){39}}\put(14,20){\line(1,0){12}}
\put(0,20){\line(-1,0){12}}
\put(-6,-5){\rotatebox[origin=c]{72}{\bigg\{}}
\put(-1,-14){$\mathfrak{1}$-$\mathfrak{2}$}
\put(16,37){\rotatebox[origin=c]{-110}{\bigg\{}}
\put(26,48){$\mathfrak{3}$-$\mathfrak{4}$}
\put(27,25){$\mathfrak{a}$}\put(43,18){$\mathfrak{b}$}
\put(12,36){$\bullet$}\put(23,-1){$\bullet$}\put(15,20){\circle{5}}\put(26,20){\circle{5}}
\put(50,-5){$\mathfrak{a}/\mathfrak{b}$: $+1$}
\put(62,35){$+2$}
\put(60,32){\line(1,0){20}}\put(60,45){\line(1,0){20}}\put(60,32){\line(0,1){13}}\put(80,32){\line(0,1){13}}
\end{picture}
\begin{picture}(95,75)(-15,-20)
\put(54,40){f)}
\put(20,20){\circle{40}}
\put(20,20){\line(0,1){20}}\put(20,20){\line(-1,-1){14}}\put(20,20){\line(1,-1){14}}
\put(1,28){\line(-2,1){10}}
\put(7,40){\rotatebox[origin=c]{-105}{\Bigg\{}}
\put(20,49){$\mathfrak{1}$-$\mathfrak{2}$}
\put(56,0){$\mathfrak{b}$}\put(8,-6){$\mathfrak{a}$}
\put(4,-16){\rotatebox[origin=c]{77}{\Bigg\{}}
\put(9,-25){$\mathfrak{3}$-$\mathfrak{4}$}
\put(20,20){\circle{5}}\put(17,28){$\bullet$}\put(3,3){$\bullet$}\put(31,3){$\bullet$}
\qbezier(20,30)(90,-10)(20,0)
\end{picture}
\end{center}
\caption{Illustration to go with the re-evaluations of the limits for finiteness set by the 4-point 3- and 4-loop one-particle irreducible loop structures. a) and b) show the 3-loop structures, while the rest show the 4-loop equivalents. Integration propagators are marked by numbers, the presence of additional legs are marked by `$+x$', either at a specific location or in general. Divisions into substructures ($\mathfrak{1}$-$\mathfrak{2}$,$\mathfrak{3}$-$\mathfrak{4}$) are indicated by dots and circles, a process which ought to be self-explanatory with the integration propagators numbered and indicated. Note that the 2-loop substructure $\mathfrak{3}$-$\mathfrak{4}$ in d) can be turned about the axis described by the dots. With the loop integration propagators moved also, configurations can be made equivalent by what will be referred to as `symmetry and renumbering'. Moreover, e) and f) are similar with respect to UV divergences, and here described in terms of the same features. For the 4-point amplitude, one more outer leg needs to be added to the propagator $\mathfrak{a}$ or $\mathfrak{b}$, in addition to two more legs, distributed at will. Note, however, that the propagator $\mathfrak{b}$ in f) describes a non-planar structure, effectively passing above one of the other propagators; it is \emph{not} describing a 4-point vertex.
\label{fig.loopDisect}}
\end{figure}
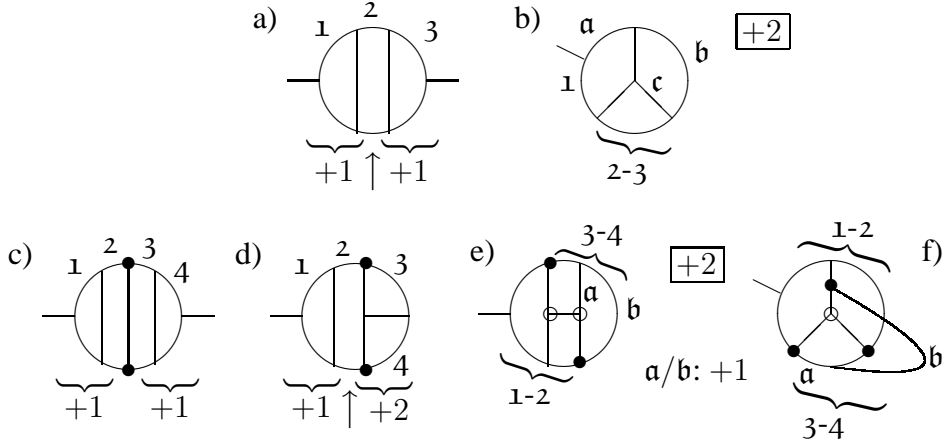

There are two structurally different 3-loop diagrams. The first structure is depicted in fig. \ref{fig.loopDisect}a); it has a 4-point configuration limited by all substructures requiring $j=4$. There are two additional legs as denoted by $+1$, but the loops ($\mathfrak{1}$,$\mathfrak{3}$) are restricted to represent 1-loop 4-point structures, contributing $(\partial^2)^0$ to the loop divergence. By that, at least $r^2$ on each outer leg in ($\mathfrak{1}$,$\mathfrak{3}$) is lost to the general regularisation, and there is at least one such on each of the two loops. In addition, the arrow marks a propagator unique to ($\mathfrak{2}$) by the distribution of the integration propagators. It cannot contribute to the divergences of ($\mathfrak{1}$,$\mathfrak{3}$), yet one derivative of the $b$ residing there is lost to ($\mathfrak{2}$) by $b^2=0$: all of the $b$ on the marked propagator cannot transfer to the loop integration propagator, giving a further loss of $r^1$. As a total of $r^5$ is lost, the entire possible divergence by \eqref{eq.nReg}: $(\partial^2)^2$ is avoided, and the limit on finiteness is $D<20/3$ by \eqref{eq.GlimitD}.

The second structure, in fig. \ref{fig.loopDisect}b), is a bit more intricate since an inner vertex is shared by all of the three loops. It is more compact, and we will see that the compact structures give rise to the worst UV divergences. Quite simply, it allows $r$ to be divided to the greatest extent between the loops, in a situation as close to \eqref{eq.PestD} as is possible to obtain. In fig. \ref{fig.loopDisect}b), it is equivalent to put one outer leg on the propagator indicated by $\mathfrak{a}$: the basic structure is completely symmetric. Furthermore, at most two outer legs can be added to the same propagator since all substructures require $j=4$ and the basic structure has $j=3$ for the separate loops. The configurations possible to distribute as such are $(2,2)$, $(2,1,1)$ or all outer legs on separate propagators. In the first scenario, the one pair is equivalently connected to $\mathfrak{a}$, at which point the second only can be placed on $\mathfrak{c}$: $\mathfrak{1}$ is a 5-point 1-loop and $\mathfrak{2}$-$\mathfrak{3}$ is a 4-point 2-loop structure. 
In the second scenario, the two last outer legs also must be added to $\mathfrak{2}$-$\mathfrak{3}$. By symmetry, it is equivalent to put the first of those on either $\mathfrak{b}$ or $\mathfrak{c}$, so that the situation either falls within the first scenario, or at least, the 4-point 2-loop by the outer leg on $\mathfrak{b}$ causes a loss of $r^2$ in comparison with \eqref{eq.nReg}. Moreover, this last observation is equally true in the third scenario, with all outer legs on different propagators, since the loop integrations can be placed on three of those with at most one outer leg shared between two loops, so that there is at least one 4-point 1-loop with an outer leg. Either way, at most $\partial^2$ can be formed, and the structure is finite in $D<6$.

In conclusion, the closer look at how $r$ might be divided yields a limit on finiteness different from \eqref{eq.PestD}, as already known from \cite{Bern2}:
\begin{equation}
D<6\qquad L=3.
\end{equation}

\subsubsection*{The 4-point $L=4$}
There are four structurally different 4-loop diagrams. The first structure is depicted in fig. \ref{fig.loopDisect}c). The configuration is limited by all substructures requiring $j=4$. There are two additional legs as denoted by $+1$, but the loops are restricted to represent a pair of 2-loop 4-point structures, visible by considering the vertices marked by dots as outer vertices, in total contributing $(\partial^2)^0$ to the loop divergence. Hence, the limit on finiteness is $D<26/4$ by \eqref{eq.GlimitD}.

The second structure, in fig. \ref{fig.loopDisect}d), similarly as for the structure $a)$ causes an $r^2$ loss by the 4-point loop $\mathfrak{1}$ and an $r^1$ loss by the loop configuration of $\mathfrak{2}$. The loops $\mathfrak{3}$-$\mathfrak{4}$ make up a 4-point 2-loop substructure giving $(\partial^2)^0$ with (by symmetry and renumbering) at most one outer leg on the propagators shared with loop $\mathfrak{2}$: three propagators, carrying a total of $r^6$, are unique to that loop structure, while excepting the integration propagators. A total of (at least) $r^9$ is lost: $(\partial^2)^3$ is lost in \eqref{eq.nReg}, so at most one such can be formed and the relevant limit is $D<6$.

The third and fourth structures in fig. \ref{fig.loopDisect}e) and f) are different in that the second is non-planar. However, it is possible to analyse them both at once. One propagator equivalently (by symmetry of the basic structure) has one leg attached as shown, in e) for the leftmost loop to be 4-point. Also in e), the rightmost loop has an outer leg attached to either one of the propagators denoted by $\mathfrak{a}$ and $\mathfrak{b}$. By symmetry and renumbering, this also goes for f), where both $\mathfrak{a}$ and $\mathfrak{b}$ can be considered to be part of a different subset of loops than the first attached leg. In total, this gives the possibility of looking at two loop structures: $\mathfrak{1}$-$\mathfrak{2}$ and $\mathfrak{3}$-$\mathfrak{4}$, sharing three propagators and two inner $R$s with at least $R^2b^3\sim r^5D^2$ shared by the two substructures, which are 4-point 2-loop structures to which two more outer legs have to be added. In fact, with $i$ shared outer legs, the shared $r$s  and $D$s amount to at least $r^{5+2i}D^{2+2i}$, by the $D$s from the added $b^i$ also effectively being shared. Anyhow, regardless of where those additional legs are placed, the shared entities cannot be counted twice, which in combination with the 4-point 2-loop structure in total corresponds to a loss of $(\partial^2)^2$ in the estimate of \eqref{eq.PestD}. Consequently, the limit on finiteness is $D<11/2$.

In conclusion, the closer look at how $r$ might be divided yields a limit on finiteness different from \eqref{eq.PestD}:
\begin{equation}\label{eq.4p4LD11.2}
D<11/2\qquad L=4,
\end{equation}
as already known from \cite{Bern3}.

\subsubsection*{Further limits on $5\leq L\leq7:$ 4- \& 5-point diagrams}
It is possible to note that all 4- and 5-point $L$-loop one-particle irreducible loop structure configurations with $L>4$ are possible to form from the 4- and 5-point 4-loop one-particle irreducible loop configurations through the addition of $L-4$ propagators connecting to the original structures. For example by starting from the 5-point 7-loop diagram, with five outer legs distributed on the inner propagators, it is possible to in an iterative manner choose a propagator with no outer leg attached, cut it and create a $(j+2)$-point $(L-1)$-loop one-particle irreducible loop structure, all the way down to $L=4$ with $j=11$. The $5$-loop structure has $j=9$ on $12$ inner propagators, making the last step feasible, and representing the last given, feasible step. Moreover, with the 5-point 7-loop given, the 4-point 6-loop naturally is accommodated for as well.

This is interesting, because $L\rightarrow L+1$ in this setting corresponds to an addition of two inner vertices and three propagators, in total $R^4b^3\sim r^{8+9}$, i.e. with $D^9$ claimed by the additional loop integration, the introduced, extra components $\sim (\partial^2)^2$. Based on the overall limit on the 4-point 4-loop given by \eqref{eq.4p4LD11.2}, \eqref{eq.GlimitD} can then be rewritten as
\begin{equation}\label{eq.do4p}
D<2+14/L \qquad 4\leq L\leq6 \quad j=4,
\end{equation}
corresponding to the results of \cite{Vanhove:2010nf,Beisert:2010jx}. The remaining question is if further $\partial^2$ can be formed in the new loop configuration.

By equivalently, at each $L\rightarrow L+1$, considering the alteration of the one-particle irreducible loop structure as consisting of two steps, a more detailed look into the situation can be provided. Firstly, two additional outer vertices are added, which by $j\rightarrow j+2$ at most can provide two new $\partial^2$. Then, all but the two outer $R$s and $D$s forced out are equivalently claimed for the formation of $(\partial^2)^2$ to be possible. After that point, the two outer legs may be considered to be connected through a new integration propagator. The added structure then describes ($D^5$,$D^6$) on the integration propagator, by ($b_0$,$b_1$), in addition to the $r$s of the $RbR$ which cannot be regularised ($r^2$,$r^3$). However, since the latter forces an additional regularisation of the $r$s on the $j$-point $(L-1)$ sub-loop structure, the calculation really is a zero-sum game: the $(\partial^2)$ can be accommodated for (in the case of $b_0$ by one $\partial^2$ being moved to the introduced loop).

On the other hand, any additional $\partial^2$ would require $\sim r^4$ to be claimed from the $(L-1)$ sub-loop structure, with $L>4$. As demonstrated during the discussion on the 4-point 4-loop structure, the base in \ref{fig.loopDisect}e) and f), used for this estimate, is highly compact and therefore contributing with the most divergent result. In addition, the structures are the most limited by shared $\sim r$s, minimal in their configurations. A saturation of the $\sim r$s possible to share have already been observed, so there is no possibility of withdrawing $r^4$ without affecting the other $\partial^2$s of the structure. The only possibility of acquiring new $\partial^2$s are by extending the c) and d) structures; a scenario by equivalence falling under the extension of e) or f), with one of the free propagators in c) and d) equivalently added. 

Hence, by the compactness of the e) and f) diagrams, the overall behaviour is limited by \eqref{eq.do4p}. Important to note is that it is merely a rough lower limit, which possibly might be further specified to something more allowing. For example, as at the transition between $L=3$ and $L=4$, the full $(\partial^2)^2$ is not always captured.
\\ \\
The 5-point setting can be analysed in the same way. There, the question of concern is a possible finiteness in $D=4$, i.e. at $L=7$, as indicated in \cite{Beisert:2010jx}. What we thereby wish to know is if $D<6$ holds for the 5-point 4-loop one-particle irreducible substructures, but only in a 5-loop equivalent sense. That is, not the actual 5-point 4-loop behaviour, but what effectively, as above for the 4-point structure, gives the 5-point 5-loop structures. In this setting, it is equivalent to consider a general 7-point version of the 4-loop structures, with any two outer vertices removed; the effective behaviour generated at the $L\rightarrow L+1$ is the same. Also, naturally, the 5-point 4-loops constructed out of the diagrams e) and f) are just as compact as the 4-point versions with respect to the formation of $\partial^2$, so with an effective $D<6$, in the equivalent setting, the overall limit would correspond to
\begin{equation}\label{eq.do5p}
D<2+16/L \qquad 5\leq L\leq7 \quad j=5.
\end{equation}
We will skip the finer points of the 5-point diagrams as well as trying to find out their strict divergences; as long as the equivalent 5-point structures correspond to $D<6$, i.e. at most $(\partial^2)^2$ formed, the worst possible divergence of the one-particle irreducible loop structures is given by \eqref{eq.do5p}.

Consider the discussion on the 4-point 4-loop one-particle irreducible loop structures right above. The 5-point versions just have one extra outer leg, and as already stated, the diagrams behave no worse than what is true for the minimal $j$, so diagram c) and d) by default fall under $D<6$. The real issue is the diagram e) and f). However, the 5-loop structures caused by an extension of f) are equivalently caused by an extension of e): any non-planar 5-loop diagram can be reduced to planar by at least six different cuts, compare e.g. to the illustration in \cite{JB}, and in the 5-point setting at least one of those is free from outer legs.

What then remains is the diagram of fig. \ref{fig.loopDisect}e) in a 5-point setting. It is possible to note that the structures with no shared outer legs at most result in the total formation of $(\partial^2)^2$. The combination of a 7-point 2-loop and a 4-point 2-loop is characterised by the latter contributing with $(\partial^2)^0$ while requiring $D^2$ from the shared $R$s, reducing the former by $\sim r^2$ to at most form $(\partial^2)^2$. The other combination of a 6-point 2-loop and a 5-point 2-loop is similarly restricted: if the latter contributes with $\partial^2$ through regularisation or free $\partial$s, the former cannot contribute with $(\partial^2)^2$.

Interestingly, in a 7-point equivalent setting, this sets the overall behaviour to $D<6$, because regardless of the distribution of 7 outer legs on the diagram of fig. \ref{fig.loopDisect}e), it can by symmetry and renumbering in combination with two outer legs equivalently being removed, present a 5-point 4-loop, divided as in e), with no shared outer legs.

\section{Conclusions \& outlook}
In this article, the pure spinor field theory results of \cite{CK2,Karlsson:2014xva} with respect to maximal supergravity have been revisited. The observation in \cite{Karlsson:2014xva} of the UV divergences only depending on one-particle irreducible loop structures constructed out of propagators and 3-point vertices in terms of certain effective operators, in total constraining the non-zero $j$-point $L$-loop structures to
\begin{gather}\begin{aligned}
j=&4:L\leq6\\
j\geq&5:L\leq7,
\end{aligned}\end{gather}
has been extended to a confirmation of the limits on the dimension for finiteness: \cite{Bern2,Bern3}
\begin{gather}\begin{aligned}
&L=1:D<8\\
2\leq&L\leq4:D<4+6/L
\end{aligned}\end{gather}
and \cite{Vanhove:2010nf,Beisert:2010jx}
\begin{equation}\label{eq.overall4}
5\leq L\leq6,\hspace{0.2cm} j=4: \quad D<2+14/L.
\end{equation}
An additional, crucial result --- possible to note for e.g. $L=7$, $j=5$ in \cite{Beisert:2010jx}, which seems to discuss the very same one-particle irreducible limit --- is
\begin{equation}\label{eq.overall5}
5\leq L\leq7, \hspace{0.2cm} j=5: \quad D<2+16/L,
\end{equation}
where both \eqref{eq.overall4} and \eqref{eq.overall5} constitute rough lower limits, possibly subject to further constraints at a more detailed analysis.

In this setting, all amplitude diagrams in maximal supergravity (where $L>7$ only is possible as a product of several one-particle irreducible structures) are concluded to be finite in
\begin{equation}
D\leq 4
\end{equation}
by the workings of the pure spinor formalism, i.e. a formulation with both on- and off-shell maximal supersymmetry. In particular, the restrictions on the UV divergences which usually are discerned in terms of U-duality, in the pure spinor formulation show in terms of the loop regularisation $r\leftrightarrow \la\lb D$ (variable/momenta equivalence) and how far those momenta can be shared within the loop structures. Furthermore, the limit on $L$ occurs not so much due to this equivalence (although partly a product thereof) as due to the insensitivity of the integration over the loop momenta to the $r\leftrightarrow \la\lb D$ conversion, thereby (through the number of $r$s present being restricted) limiting $L$. What this corresponds to in terms of other approaches to the UV divergences in maximal supergravity is, however, difficult to tell, although it would be highly interesting to see a corresponding analysis in a different setting.

The result is intriguing with respect to the UV finiteness of maximal supergravity in four dimensions. This is a scenario traditionally regarded as highly unlikely. Still, the analysis in a pure spinor field theory setting indicate precisely that. Admittedly, the pure spinor formalism is difficult to interpret in terms of the component fields of ordinary maximal supergravity, but despite sometimes being regarded as somewhat obscure, it encodes the same physics. In total, a confirmation of the results in a different setting would be most welcome. There is a possibility of an overall behaviour of maximal supergravity (disregarding the specific dependence on $L$ ) just as in maximally supersymmetric Yang--Mills theory: UV finiteness in $D=4$.

\section*{Ackowledgements}
I would like to thank M. Cederwall for helpful discussions.

\appendix
\section{Properties of the pure spinor}\label{app.Lb}
The spinors in $D=11$ supergravity are symplectic. To (in the flat setting of the pure spinor formalism) capture the relevance of ordering, all spinor indices are chosen to be lower and contracted through $\varepsilon^{\alpha\beta}=\varepsilon^{[\alpha\beta]}$: $(\la\g_a)_\gamma= \la_\alpha\varepsilon^{\alpha\beta}(\g_a)_{\beta\gamma}$.

The Fierz identity is
\begin{equation}
(AB)(CD)=\sum\limits_{p=0}^5
\frac{1}{32p!}(C\g^{a_1\ldots a_p}B)(A\g_{a_p\ldots a_1}D),
\end{equation}
assuming bosonic entities $(A,B,C,D)$; if the statistics differ from this, the addition of an appropriate sign suffices for the correct expression to be obtained. In the presence of two pure spinors, this furthermore reduces to
\begin{equation}
(A\la)(\la B)=-\frac{1}{64}(\la\g^{ab}\la)(A\g_{ab}B)
+\frac{1}{3840}(\la\g^{abcde}\la)(A\g_{abcde}B).
\end{equation}
In total, there are several useful identities for the pure spinor, compare to \cite{CK2} and \cite{Karlsson:2014xva}, not the least in the presence of $L^{(p)}$. For example,
\begin{gather}\begin{aligned}\label{eq.appContr}
(\g_j\lb)_\alpha(\lb\g^{ij}\lb)&=0\\
(\lb\g^{[ij}\lb)(\lb\g^{kl]}r)&=0\\
(\lb\g^i{}_k\lb)(\lb\g^{jk}r)&=(\lb\g^{ij}\lb)(\lb r)\\
(\lb\g^i{}_kr)(\lb\g^{jk}r)&=(\lb\g^{ij}r)(\lb r)+\frac{1}{2}(\lb\g^{ij}\lb)(rr)\\
(\lb\g^{[ab}\lb)(\lb\g^{c]d}\lb)&=0\\
(\lb\g^{ab}\lb)(\lb\g^{cd}\lb)f_{ac}&=\frac{1}{2}(\lb\g^{ac}\lb)(\lb\g^{bd}\lb)f_{ac}\\
L^{(1)}_{ab,cd}f^{abc}&=(\lb\g_{ab}\lb)(\lb\g_{cd}r)f^{abc}.
\end{aligned}\end{gather}

\def\prp {Phys. Rept.}
\def\prv{Phys. Rev.}
\def\prl{Phys. Rev. Lett.}
\def\prva{Phys. Rev. A}
\def\prvd{Phys. Rev. D}
\def\jhep{\mbox{J. High} Energy Phys.}
\def\pLondA{Proc. Roy. Soc. Lond. A}
\def\cqg{Class. Quantum Grav.}
{\small
\bibliography{references.bib}
\bibliographystyle{JHEP1}
}
\end{document}